\documentclass[aps,prb,showpacs,superscriptaddress,twocolumn]{revtex4}
\usepackage{amsmath}
\usepackage{graphicx}

\begin{document}

\title{Probing Nano-Mechanical QED Effects}
\author{Y. B. Gao}
\email{ybgao@bjut.edu.cn} \affiliation{Advanced Science Institute,
The Institute of Physical and Chemical Research (RIKEN), Wako-shi
351-0198, Japan} \affiliation{College of Applied Sciences, Beijing
University of Technology, Beijing, 100124, China}
\author{S. Yang}
\affiliation{Advanced Science Institute, The Institute of Physical and Chemical Research
(RIKEN), Wako-shi 351-0198, Japan}
\affiliation{Institute of Theoretical Physics, The Chinese Academy of Sciences, Beijing,
100190, China}
\author{Yu-xi Liu}
\affiliation{Advanced Science Institute, The Institute of Physical and Chemical Research
(RIKEN), Wako-shi 351-0198, Japan}
\affiliation{CREST, Japan Science and Technology Agency (JST), Kawaguchi, Saitama
332-0012, Japan}
\author{C. P. Sun}
\affiliation{Advanced Science Institute, The Institute of Physical and Chemical Research
(RIKEN), Wako-shi 351-0198, Japan}
\affiliation{Institute of Theoretical Physics, The Chinese Academy of Sciences, Beijing,
100190, China}
\author{Franco Nori}
\affiliation{Advanced Science Institute, The Institute of Physical and Chemical Research
(RIKEN), Wako-shi 351-0198, Japan}
\affiliation{CREST, Japan Science and Technology Agency (JST), Kawaguchi, Saitama
332-0012, Japan}
\affiliation{Center for Theoretical Physics, Physics Department, Center for the Study of
Complex Systems, The University of Michigan, Ann Arbor, Michigan 48109-1040,
USA}
\date{\today }

\begin{abstract}
We propose and study an ``intrinsic probing'' approach, without introducing
any external detector, to mimic cavity QED effects in a qubit-nanomechanical
resonator system. This metallic nanomechanical resonator can act as an
intrinsic detector when a weak driving current passes through it. The
nanomechanical resonator acts as both the cavity and the detector. A cavity
QED-like effect is demonstrated by the correlation spectrum of the
electromotive force between the two ends of the nanomechanical resonator.
Using the quantum regression theorem and perturbation theory, we
analytically calculate the correlation spectrum. In the weak driving limit,
we study the effect on the vacuum Rabi splitting of both the strength of the
driving as well as the frequency-detuning between the charge qubit and the
nanomechanical resonator. Numerical calculations confirm the validity of our
intrinsic probing approach.
\end{abstract}

\maketitle

\section{introduction}

Recently, nanomechanical resonators (NAMRs) are attracting
considerable attention (see, e.g.,
Refs.~\onlinecite{Cleland02,Blencowe04,Blencowe07,Mahboob08}). Also,
mechanical analogues of cavity QED have been theoretically studied
in coupled systems between nanomechanical resonators and superconducting qubits (see, e.g., Refs.~\onlinecite%
{Schwab02,Schwab03,Cleland04,Wang04,Nori07}). Various effects in these
nanomechanical QED systems were investigated, including: quantum
measurements~\cite{Schwab03}, the quantum squeezing of the NAMRs~\cite%
{Wang04, Xue07-1, Xue07}, and the cooling of the NAMRs~\cite%
{Martin04,Zhang05,Schwab06,Nori08,Liyong08}. Some of these
theoretical proposals have recently become experimentally testable
due to the recent advances in NAMRs and superconducting qubits.
Numerous Josephson-junction-based superconducting qubits have been
experimentally realized (see, e.g., the reviews~\cite{Makhlin01,
You05, Clarke08,Wendin06}), while studies on NAMRs with vibration
frequencies of the order of a GHz are approaching the quantum
regime.

References \cite{Lambert08,Ouyang08} recently studied a NAMR coupled
to a double-quantum dot. In Ref.~\onlinecite{Lambert08}, the
spectrum of the transport current was used to study the quantum
behavior of this system. The electron transport through a mobile
island (i.e., a nanomechanical oscillator) with two energy levels
was studied in Ref.~\onlinecite{Johansson08}. There, the qubit was
embedded in the NAMR.

To study cavity QED analogues in a NAMR-qubit system, a crucial issue is how
to make the quantum measurement on this coupled system. Quantum measurements
involve subtle interactions between the system and the detector. To carry
out a quantum measurement, an external probing instrument is typically
coupled to the measured system. Examples of this include: a single electron
transistor~\cite{Shnirman98} coupled to a charge qubit, a transmission line
resonator~\cite{Blais04,You03,Liu04} coupled to a charge qubit, or a shunted
dc-SQUID coupled to a flux qubit (see, e.g., Refs.~\onlinecite%
{Mooij99,ilichev,shnirman}). In general, these coupled systems can
be modeled by the Jaynes-Cummings Hamiltonian and demonstrate
several analogues to cavity QED effects~\cite{Raimond01,Girvin08},
such as vacuum Rabi splitting. These effects can be used to verify
the coherent coupling between a superconducting qubit and a
measuring device.

This study is mainly motivated by recent experiments on a
high-frequency metallic NAMR~\cite{Litf08}. Previously, non-metallic
NAMRs were often studied and therefore no efficient current would
pass through these non-metallic NAMRs, and thus no mechanical force
acting on the NAMRs could be induced to implement quantum
measurements. In this case, an external instrument needs to be
integrated to probe the coupling between the NAMR and the qubit.
Here, we study how to probe a cavity QED analogue for a metallic
NAMR~\cite{Litf08} coupled to a superconducting qubit \emph{without}
introducing an external detector.

In this proposal, the information on the coherent coupling between
the superconducting charge qubit and the metallic NAMR can be read
out by measuring the induced electromotive force between the two
ends of the NAMR. This electromotive force is generated by a current
passing through a metallic NAMR in which a magnetic field is
applied. There are at least two advantages for this intrinsic
probing mechanism: (i) the coupling between the metallic NAMR and
the qubit can be turned on or off by the externally-applied voltage,
and then the information can be read out in a controllable way; (ii)
no external probing instrument needs to be introduced, in contrast
to the proposal in Ref.~\onlinecite{Wei06}.

This paper is organized as follows. In Sec.~II, we describe the proposed
model, and write the Hamiltonian for a charge qubit interacting with a
driven metallic NAMR. In Sec.~III, we calculate the spectrum of the two-time
correlation function for the induced electromotive force using the quantum
regression theorem~\cite{Scully97} and perturbation theory. In the weak
driving limit, we study how the Rabi splitting depends on both the strength
of the driving current which passes through the NAMR as well as the detuning
between the frequencies of the charge qubit and the nanomechanical
resonator. Using numerical calculations, we demonstrate that our analytical
results are valid in the weak driving limit. Finally, we summarize our
conclusions.

\section{model}

\begin{figure}[th]
\includegraphics[bb=29 130 563 743,width=8cm,clip]{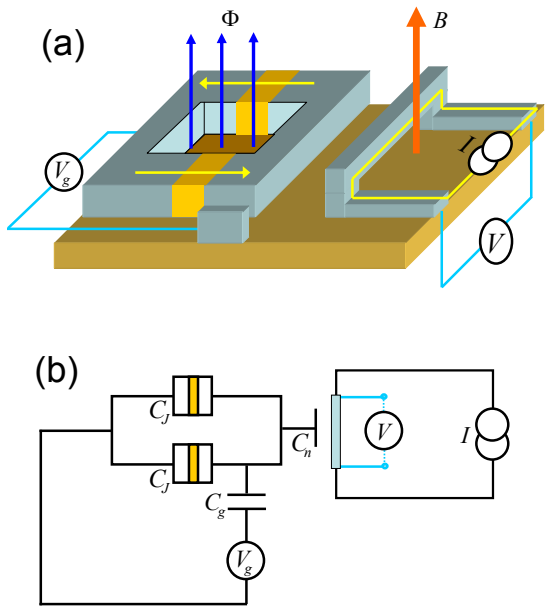}
\caption{(Color online) (a) Schematic diagram of a charge qubit
(grey loop on the left) capacitively coupled to a metallic
nanomechanical resonator (NAMR) shown on the right. An external ac
current, $I(t)=I_{0}\sin(\protect\omega_{p} t)$, shown in yellow,
passes through the NAMR. Also a magnetic field $B$ (red line with
the up arrow) is applied to the NAMR. The magnetic flux $\Phi$
through the SQUID is denoted by the blue
upward-pointing arrows. The induced electromotive force $V$ (defined in Eq.~(\protect\ref%
{eq:1})) at both ends of the NAMR can be used to detect information on the
quantum-coherent coupling. A simple circuit diagram for (a) is given in (b).
Here, $C_{J}$ and $C_{g}$ represent the capacitances for the Josephson
junctions and the gate capacitor, respectively. Also, $V_{g}$ is the gate
voltage applied to the qubit via the gate capacitor.}
\label{setup}
\end{figure}

As shown in Fig.~\ref{setup}, we study a metallic NAMR, which is
capacitively coupled to a SQUID-based Cooper pair box (qubit). The
distributed capacitance between the NAMR and the superconducting island of
the qubit is denoted by $C_{n}(x)$. Here no voltage is applied to this
distributed capacitor, in contrast to previous work~\cite{Wang04}. The
electromagnetic force drives the NAMR to oscillate and the induced
electromotive force between the two ends of the NAMR can be described by
\begin{equation}
V=Bl\;\frac{\text{d} x}{\text{d} t}=Bl\frac{p}{M}.  \label{eq:1}
\end{equation}
Here $l$ and $M$ denote the length and the mass of the NAMR, respectively.
Also, $(\text{d}x/\text{d}t)=(p/M)$ denotes the velocity of the NAMR; $p$
represents the momentum for the center of mass of the NAMR; $x$ denotes the
small displacement of the NAMR around the coordinate-axis origin, and $x=0$
when no ac current passes through the NAMR. The Hamiltonian $H_{\mathrm{NAMR}%
}$ for the ac-current-driven NAMR is given by
\begin{equation}
H_{\mathrm{NAMR}}=\frac{p^{2}}{2M}+\frac{1}{2}M\Omega ^{2}x^{2}-lBI(t)x,
\label{Ham-namr}
\end{equation}
where the canonical coordinate $x$ and momentum $p$ for the NAMR are assumed
to satisfy the commutation relation $\left[ x,p\right] =i$. Hereafter, we
assume $\hbar =1$. The parameter $\Omega$ denotes the oscillating frequency
of the NAMR. The $x$ and $p$ of the NAMR can be represented by the
annihilation $a$ and creation $a^{\dag }$ operators as,
\begin{eqnarray}
x&=&\frac{1}{\sqrt{2M\Omega }}\left( a^{\dag}+a\right) ,  \label{eq:2-1} \\
p&=&i\sqrt{\frac{M\Omega}{2}}\left(a^{\dag }-a\right).  \label{eq:2-2}
\end{eqnarray}%
Thus the Hamiltonian in Eq.~(\ref{Ham-namr}) can be rewritten as
\begin{equation}
H_{\mathrm{NAMR}}=\Omega\, a^{\dagger} a- \frac{lB}{\sqrt{2M\Omega }}\left(
a+a^{\dag }\right) I(t).  \label{eq:3}
\end{equation}

Let us now assume that the SQUID has two identical Josephson junctions, each
with the Josephson energy $E_{J0}$ and capacitance $C_{J}$. A control gate
voltage $V_{g}$ is applied to the Cooper-pair box via the gate capacitor
with the capacitance $C_{g}$. The Hamiltonian of the box can be written~\cite%
{You05,Clarke08,Makhlin99} as
\begin{equation}
H_{c}=\frac{2e^{2}}{C_{\Sigma }(x) }\left( n-n_{g} \right) ^{2}+E_{J}\cos
\varphi  \label{Ham-JJ}
\end{equation}%
with the controllable effective Josephson energy $E_{J}=2E_{J0}\cos (\pi
\Phi/\Phi_{0})$ of the SQUID. Here, $\Phi$ is the magnetic flux through the
SQUID loop and $\Phi_{0}$ is the flux quantum. The total capacitance $%
C_{\Sigma }(x)$ connected to the superconducting island is given by
\begin{equation*}
C_{\Sigma }(x) =2 C_{J}+C_{g}+C_{n}(x)\, .
\end{equation*}%
The effective Cooper pair number $n_{g}$ in the superconducting island is $%
n_{g} =C_{g}V_{g}/(2e)$. We assume that the distance $d$ between the NAMR
and the superconducting island is much larger than the amplitude $x$ of the
oscillation of the NAMR, i.e., $d\gg x$. In this case, the distributed
capacitance can be approximately written as
\begin{equation}
C_{n}(x) \simeq C_{n}\left( 1-\frac{x}{d}\right)
\end{equation}
to first order in $x/d$. The island's charging energy is $E_{C}=e^{2}/\left[
2C_{\Sigma }(0) \right] $ and $n_{g}$ is near the degeneracy point $1/2$. In
these conditions, the box can be reduced to a two-level quantum system and
the Hamiltonian in Eq.~(\ref{Ham-JJ}) can be reduced to
\begin{equation}
H_{c}=4E_{C}\left( n_{g}-\frac{1}{2}\right) \left( 1+\frac{C_{n}}{C_{\Sigma
}(0) }\frac{x}{d}\right) \bar{\sigma}_{z}-\frac{1}{2}E_{J}\bar{\sigma}_{x}
\label{Ham-c1}
\end{equation}
in spin-$1/2$ notation with the quasi-spin operators
\begin{eqnarray*}
\bar{\sigma}_{z} &=&\left\vert 0\right\rangle _{cc}\left\langle 0\right\vert
-\left\vert 1\right\rangle _{cc}\left\langle 1\right\vert , \\
\bar{\sigma} _{x}&=&\left\vert 0\right\rangle _{cc}\left\langle 1\right\vert
+\left\vert 1\right\rangle_{cc}\left\langle 0\right\vert ,
\end{eqnarray*}%
which are defined in the basis of the charge states $\left\vert
0\right\rangle _{c}$ and $\left\vert 1\right\rangle _{c}$. Equation~(\ref%
{Ham-c1}) is used to describe the interaction between the NAMR and the
charge qubit.

We now reconstruct a set of spin operators $\{\sigma _{z},$ $\sigma _{\pm }\}
$ with
\begin{equation}
\sigma _{+}\left\vert 0\right\rangle =\left\vert 1\right\rangle ,\ \ \sigma
_{-}\left\vert 1\right\rangle =\left\vert 0\right\rangle ,
\end{equation}%
\label{eq:6} and
\begin{eqnarray}
\left\vert 1\right\rangle &=&\cos \frac{\theta }{2}\left\vert 0\right\rangle
_{c}-\sin \frac{\theta }{2}\left\vert 1\right\rangle _{c},  \label{eq:7} \\
\left\vert 0\right\rangle &=&\sin \frac{\theta }{2}\left\vert 0\right\rangle
_{c}+\cos \frac{\theta }{2}\left\vert 1\right\rangle _{c} ,  \label{eq:8}
\end{eqnarray}
where the mixing angle $\theta$ is determined by
\begin{equation*}
\tan \theta =\frac{E_{J}}{4E_{C}\left( 2n_{g}-1\right)}.
\end{equation*}
In the new basis $|0\rangle$ and $|1\rangle$, the Hamiltonian in Eq.~(\ref%
{Ham-c1}) becomes
\begin{equation}
H_{c}=\frac{\omega _{a}}{2}\sigma _{z}+\frac{4E_{C}\;C_{n}}{d\,C_{\Sigma
}(0) }\left(n_{g}-\frac{1}{2}\right) \left( \cos \theta \sigma _{z}+\sin
\theta \sigma _{x}\right) x  \label{Ham-c2}
\end{equation}
with the qubit frequency
\begin{equation}
\omega _{a}=\sqrt{16E_{C}^{2}\left( 2n_{g}-1\right) ^{2}+E_{J}^{2}}.
\end{equation}
Notice that the coupling between the coordinate $x$ of the NAMR and
the qubit has a term proportional to the gate voltage ($\propto
n_{g}$). Thus, the gate voltage $V_{g}$ can control this coupling.
Using Eq.~(\ref{Ham-c2}) and also considering the driven NAMR, we
can now write down the total Hamiltonian of the driven NAMR
interacting with the charge qubit
\begin{equation}
H=H_{c}+H_{\mathrm{NAMR}}.  \label{eq:9}
\end{equation}
Here, the Hamiltonians $H_{\mathrm{NAMR}}$ and $H_{c}$ are given by Eqs.~(%
\ref{eq:3}) and (\ref{Ham-c2}), respectively.

In the rotating reference frame at the driven frequency $\omega _{p}$, for
both the qubit and the NAMR, through the unitary transformation
\begin{equation}
U=\exp\left[-i\omega_{p}\left(\frac{1}{2}\sigma_{z}-a^{\dagger}a\right)t%
\right],
\end{equation}
the total Hamiltonian in Eq.~(\ref{eq:9}) is converted into an effective
time-independent Hamiltonian
\begin{equation}
H_{\mathrm{eff}}=\Delta _{a}\sigma _{+}\sigma _{-}+g\left( a\sigma
_{+}+a^{\dag }\sigma _{-}\right) +\Delta a^{\dag }a-\xi \left( a+a^{\dag
}\right) .  \label{Ham}
\end{equation}
Here, the fast oscillating terms $\exp (i\omega _{d}t)$ and $\exp (2i\omega
_{d}t)$ have been neglected, and we also used the identity, $\sigma
_{z}=2\sigma _{+}\sigma _{-}-I$, where $I$ is the unit operator. In Eq.~(\ref%
{Ham}), the detuning $\Delta _{a}$ between the frequencies of the qubit and
the ac driving current is
\begin{equation}
\Delta _{a}\ \equiv \ \Delta _{\mathrm{qubit-current}}\ =\ \omega
_{a}-\omega _{p}\, .
\end{equation}
The detuning $\Delta$ between the frequencies of the NAMR and the ac driving
current is
\begin{equation}
\Delta\ \equiv\ \Delta_{\mathrm{NAMR-current}} \ =\ \Omega -\omega _{p}\, .
\end{equation}
The interaction strength $g$ (between the qubit and the NAMR) is
\begin{equation*}
g\ \equiv\ g_{\mathrm{qubit-NAMR}}\ =\ \left( n_{g}-\frac{1}{2}\right) \frac{%
1}{\sqrt{2M\Omega }}\frac{4E_{C}\; C_{n}}{d\,C_{\Sigma }( 0) }\sin \theta \,,
\end{equation*}
which can be switched off when $n_{g}=1/2$. The coupling strength (between
the NAMR and the ac driving current) is
\begin{equation}
\xi\ \equiv\ \xi_{\mathrm{NAMR-current}} \ =\ \frac{lBI_{0}}{2}\frac{1}{%
\sqrt{2M\Omega }}.  \label{xi}
\end{equation}
Note that the coupling $g$ is proportional to the gate voltage, while the
other coupling strength $\xi$ is proportional to $B I_{0}$.

\section{correlation spectrum of the induced electromotive force}

The first three terms of the right hand side of $H_{\mathrm{eff}}$ in Eq.~(%
\ref{Ham}) describe the Jaynes-Cummings Hamiltonian, which was extensively
studied in cavity QED. This QED analogue of the qubit-NAMR, described in
Eq.~(\ref{Ham}), can be studied via the correlation spectrum $S_{V}(\omega)$
of the induced electromotive force
\begin{equation}
V=iBl\sqrt{\frac{\Omega}{2M}}(a^\dagger-a),  \label{eq:20}
\end{equation}
which is obtained from Eq.~(\ref{eq:1}) by replacing the momentum operator $p$ with Eq.~(\ref%
{eq:2-2}). The correlation spectrum $S_{V}(\omega)$ of the induced
electromotive force $V$ can be calculated via
\begin{equation}
S_{V}( \omega) =\frac{1}{\pi }\text{Re}\int_{0}^{\infty }d\tau e^{i\omega
\tau }\left\langle V(0) V(\tau) \right\rangle .  \label{eq:21}
\end{equation}
Equation~(\ref{eq:20}) shows that the two-time correlation function $%
\left\langle V(0) V(\tau) \right\rangle$ in Eq.~(\ref{eq:21}) can be
calculated as
\begin{eqnarray}
\left\langle V(0) V(\tau) \right\rangle &\propto &\left\langle a(0) a^{\dag
}(\tau) \right\rangle +\left\langle a^{\dag }(0) a(\tau) \right\rangle
\notag \\
&&-\left\langle a(0) a(\tau) \right\rangle -\left\langle a^{\dag }(0)
a^{\dag }(\tau) \right\rangle .  \label{correlation}
\end{eqnarray}

\subsection{Master equation and solutions}

To obtain the correlation spectrum, we start from the master equation~\cite%
{book} of the reduced density matrix $\rho$ for the qubit-NAMR system
\begin{eqnarray}
\dot{\rho} &=&-i\left[ H_{\mathrm{eff}},\rho \right] +\kappa \left( 2a\rho
a^{\dag }-a^{\dag }a\rho -\rho a^{\dag }a\right) +  \notag \\
&&\frac{\gamma }{2}\left( 2\sigma _{-}\rho \sigma _{+}-\sigma _{+}\sigma
_{-}\rho -\rho \sigma _{+}\sigma _{-}\right) ,  \label{master equation}
\end{eqnarray}
where the latter two terms describe the decays of the NAMR and the charge
qubit, respectively. The parameters $\kappa$ and $\gamma$ denote the decay
rates of the NAMR and the qubit, respectively. We also use the Markov
approximation when Eq.~(\ref{master equation}) is derived. For convenience
below, we now define the number operator, $N=\sigma_{+}\sigma _{-}+a^{\dag }a
$, to characterize the total excitation of the qubit-NAMR. Obviously, $N$
satisfies
\begin{equation*}
N\left\vert j,k\right\rangle =\left( j+k\right) \left\vert j,k\right\rangle .
\end{equation*}
Here, the index $j$ represents the states of the charge qubit. When the
qubit is in an excited state, we take $j=1$, otherwise $j=0$. Also $k$
denotes the phonon number of the oscillating NAMR, i.e., $%
a^{\dagger}a|k\rangle=k|k\rangle$.

We are only interested in the weak driving limit, i.e.,
\begin{equation*}
\xi\ \equiv\ \xi_{\mathrm{NAMR-current}} \ \rightarrow \ 0.
\end{equation*}
In this limit, we only need to consider the zero- and one-particle
excitations; then $N$ satisfies the condition
\begin{equation}
N=j+k=0,\,\, 1.
\end{equation}
The Hilbert space for the reduced density matrix is now limited to a smaller
subspace with a truncated basis
\begin{equation}
\left\{ \left\vert j,k\right\rangle ,\,\, j+k=0,\,\,1\right\} .
\label{eq:25}
\end{equation}
Therefore, in this truncated basis, the density matrix elements satisfy the
following equations
\begin{eqnarray}
\frac{d\rho _{00,00}}{d\tau } &=&i\xi \rho _{01,00}-i\xi \rho
_{00,01}+2\kappa \rho _{01,01}+\gamma \rho _{10,10},  \notag \\
\frac{d\rho _{00,01}}{d\tau } &=&\left( i\Delta -\kappa \right) \rho
_{00,01}+ig\rho _{00,10}  \notag \\
&&+i\xi \left( \rho _{01,01}-\rho _{00,00}\right) ,  \notag \\
\frac{d\rho _{00,10}}{d\tau } &=&\left( i\Delta _{a}-\frac{\gamma }{2}%
\right) \rho _{00,10}+ig\rho _{00,01}+i\xi \rho _{01,10},  \notag \\
\frac{d\rho _{01,01}}{d\tau } &=&-2\kappa \rho _{01,01}+ig\left( \rho
_{01,10}-\rho _{10,01}\right)  \notag \\
&&+i\xi \left( \rho _{00,01}-\rho _{01,00}\right) ,  \notag \\
\frac{d\rho _{01,10}}{d\tau } &=&\left( i\Delta _{a}-i\Delta -\kappa -\frac{%
\gamma }{2}\right) \rho _{01,10}  \notag \\
&&+ig\left( \rho _{01,01}-\rho _{10,10}\right) +i\xi \rho _{00,10},  \notag
\\
\frac{d\rho _{10,10}}{d\tau } &=&-\gamma \rho _{10,10}+ig\left( \rho
_{10,01}-\rho _{01,10}\right).  \label{elements}
\end{eqnarray}%
The other non-diagonal matrix elements $\rho_{01,00}(\tau)$, $%
\rho_{10,00}(\tau)$, and $\rho_{10,01}(\tau)$ can be easily obtained by
taking the complex conjugates of $\rho_{00,01}(\tau)$, $\rho_{00,10}(\tau)$,
and $\rho_{01,10}(\tau)$, e.g., $\rho_{01,00}(\tau)= \left[ \rho
_{00,01}(\tau) \right] ^{\ast }$ when their solutions, e.g., $\rho
_{00,01}(\tau)$, are obtained using Eq.~(\ref{elements}).

In the weak driving limit, $\xi \ll g$, we take the population in the ground
state as $\rho _{00,00}=1$. We also find that two diagonal matrix elements ($%
\rho _{10,10}$ and $\rho _{01,01}$) and two off-diagonal matrix elements ($%
\rho _{01,10}$ and $\rho _{10,01}$) are proportional to $\xi ^{2}$. The
other ones are proportional to $\xi $. Using perturbation theory, we only
keep the terms to first order in $\xi$ for the reduced matrix elements in
Eq.~(\ref{elements}) and then we can obtain
\begin{eqnarray}
\dot{\rho}_{00,01} &=&\left( i\Delta -\kappa \right) \rho _{00,01}+ig\rho
_{00,10}-i\xi ,  \label{eq:26} \\
\dot{\rho}_{00,10} &=&\left( i\Delta _{a}-\frac{\gamma }{2}\right) \rho
_{00,10}+ig\rho _{00,01}.  \label{eq:27}
\end{eqnarray}%
Applying the Laplace transformation to Eq.~(\ref{eq:26}) and Eq.~(\ref{eq:27}%
), the solutions of the matrix element $\rho _{00,01}(\tau) $ can be easily
obtained as
\begin{equation}
\rho _{00,01}(\tau) =\eta _{12}e^{-\lambda _{1}\tau }+\eta _{21}e^{-\lambda
_{2}\tau }+\varepsilon ,  \label{eq:28}
\end{equation}%
with parameters
\begin{eqnarray*}
\varepsilon &=&\frac{i\xi \left( i\Delta _{a}-\frac{\gamma }{2}\right) }{%
\lambda _{1}\lambda _{2}}, \\
\eta _{mn}&=&\mu _{mn}\rho _{00,01}(0) +\chi _{mn}\rho _{00,10}( 0) +i\xi
\nu _{mn}.
\end{eqnarray*}%
Other parameters $\mu_{mn}$, $\chi_{mn}$, $\nu_{mn}$, $\lambda_{1}$, and $%
\lambda_{2}$ in Eq.~(\ref{eq:28}) are
\begin{eqnarray}
\mu _{mn} &=&\frac{\lambda _{m}+\left( i\Delta _{a}-\frac{\gamma }{2}\right)
}{\left( \lambda _{m}-\lambda _{n}\right) },  \notag \\
\chi _{mn} &=&\frac{\lambda _{m}+\left( i\Delta _{a}-\frac{\gamma }{2}%
\right) }{\lambda _{m}\left( \lambda _{m}-\lambda _{n}\right) },  \notag \\
\nu_{mn} &=&\frac{g}{i\left( \lambda _{m}-\lambda _{n}\right) },
\label{parameter-2}
\end{eqnarray}%
and%
\begin{equation*}
\lambda _{m}=\Gamma +\frac{i}{2}\left[\left( -1\right) ^{m}\sqrt{\left(
\delta -i\left( \kappa -\frac{\gamma }{2}\right) \right) ^{2}+4g^{2}}-\left(
\Delta _{a}+\Delta \right)\right] .
\end{equation*}%
for $m\left( \neq n\right) =1,\,2$. Where we define the frequency detuning
\begin{equation}
\delta =\Delta _{a}-\Delta =\omega _{a}-\Omega,  \label{delta}
\end{equation}
and the parameter $\Gamma$ is given by
\begin{equation}
\Gamma = \frac{\kappa}{2} +\frac{\gamma}{4}.
\end{equation}

The parameters $\lambda _{m}$ can be further expressed as, $\lambda
_{m}=\Gamma _{m}+i\varphi _{m}$, with real part
\begin{equation*}
\Gamma _{m}=\Gamma +\frac{1}{2}\left( -1\right) ^{m}\left(
a^{2}+b^{2}\right) ^{\frac{1}{4}}\sin \left[ \frac{1}{2}\arctan \left( \frac{%
b}{a}\right) \right] ,
\end{equation*}%
and imaginary part
\begin{equation*}
\varphi _{m}=\frac{1}{2}\left\{ \left( -1\right) ^{m}\left(
a^{2}+b^{2}\right) ^{\frac{1}{4}}\cos \left[ \frac{1}{2}\arctan \left( \frac{%
b}{a}\right) \right] -\left( \Delta _{a}+\Delta \right) \right\} .
\end{equation*}%
Here, the parameters $a$ and $b$ are
\begin{eqnarray*}
a &=&\delta ^{2}-\left( \kappa -\frac{\gamma }{2}\right) ^{2}+4g^{2}, \\
b &=&2\delta \left( \kappa -\frac{\gamma }{2}\right) .
\end{eqnarray*}

\subsection{Correlation spectrum}

The correlation function, e.g., $\left\langle a^{\dag }(0) a(\tau)
\right\rangle$, is given by
\begin{equation}
\left\langle a^{\dag }(0) a(\tau) \right\rangle =\text{Tr}\left\{ a( 0)
A(\tau) \right\}  \label{eq:33-1}
\end{equation}%
with
\begin{equation}
A( \tau ) =U( \tau) \rho (0) a^{\dag }(0) U^{\dag }(\tau).
\end{equation}
Using the quantum regression theorem~\cite{Scully97}, the correlation
function in Eq.~(\ref{eq:33-1}) can be written as
\begin{equation}
\left\langle a^{\dag }(0) a(\tau) \right\rangle =A_{01,00}(\tau ) ,
\end{equation}
with
\begin{equation}
A_{01,00}(\tau)=H_{A,12}^{\ast }e^{-\lambda _{1}^{\ast }\tau
}+H_{A,21}^{\ast }e^{-\lambda _{2}^{\ast }\tau }+\varepsilon ^{\ast }.
\label{eq:33}
\end{equation}
Here, the initial operator $A(0)$ is assumed to be
\begin{equation}
A(0)=\rho^{ss}a^{\dagger}(0)
\end{equation}
when we calculate the time-dependent matrix element $A_{01,00}(\tau)$ in
Eq.~(\ref{eq:33}). The ``$ss$" in the superscript of the reduced density
matrix $\rho$ denotes the ``steady state''. The parameters $H_{A,12}$ and $%
H_{A,21}$ in Eq.~(\ref{eq:33}) are expressed as
\begin{equation}
H_{A,mn} =\mu^{*}_{mn}A_{01,00}(0) +\chi ^{*}_{mn}A_{10,00}(0) -i\xi
\nu^{*}_{mn},  \label{eq:37}
\end{equation}
with the subscript either $mn=12$ or $mn=21$. The parameters $A_{00,01}(0)$
and $A_{00,10}(0)$ denote the matrix elements of the operator $A(0)$ in the
truncated basis defined in Eq.~(\ref{eq:25}), e.g.,
\begin{equation}
A_{00,01}(0)=\langle 00|A(0)|01\rangle=\langle
00|\rho^{ss}a^{\dagger}(0)|01\rangle.  \label{eq:39-1}
\end{equation}
These matrix elements $A_{01,00}(0)$ and $A_{10,00}(0)$ can be
straightforwardly obtained as
\begin{eqnarray}
A_{01,00}(0) &=&\rho_{01,01}^{ss}, \\
A_{10,00}(0) &=&\rho _{10,01}^{ss},
\end{eqnarray}
where $\rho_{01,01}^{ss}$ and $\rho _{10,01}^{ss}$ denote the
``steady-state'' matrix elements of the reduced density matrix $\rho $.

Using the same procedure, other correlation functions can also be obtained
as
\begin{eqnarray}
\left\langle a(0) a^{\dag }(\tau) \right\rangle &=&\text{Tr}\left\{ a^{\dag
}(0) B(\tau) \right\} =B_{00,01}(\tau) ,  \label{eq:38} \\
\left\langle a(0) a(\tau) \right\rangle &=&\text{Tr}\left\{ a( 0) C( \tau)
\right\} =C_{01,00}( \tau) ,  \label{eq:39} \\
\left\langle a^{\dag }(0) a^{\dag }(\tau) \right\rangle &=&\text{Tr}\left\{
a^{\dag }(0) D(\tau) \right\} =D_{00,01}(\tau ),  \label{eq:40}
\end{eqnarray}
with
\begin{eqnarray}
B(\tau) &=&U(\tau) \rho (0) a(0) U^{\dag }(\tau) ,  \label{eq:41} \\
C(\tau) &=&U(\tau) \rho (0) a(0) U^{\dag }(\tau),  \label{eq:42} \\
D(\tau) &=&U(\tau) \rho (0) a^{\dag }(0) U^{\dag }(\tau ),  \label{eq:43}
\end{eqnarray}%
and
\begin{eqnarray}
B_{00,01}(\tau) &=&H_{B,12}e^{-\lambda _{1}\tau }+H_{B,21}e^{-\lambda
_{2}\tau }+\varepsilon ,  \label{eq:44} \\
C_{01,00}(\tau) &=&H_{C,12}^{\ast }e^{-\lambda _{1}^{\ast }\tau
}+H_{C,21}^{\ast }e^{-\lambda _{2}^{\ast }\tau }+\varepsilon ^{\ast },
\label{eq:45} \\
D_{00,01}(\tau) &=&H_{D,12}e^{-\lambda _{1}\tau }+H_{D,21}e^{-\lambda
_{2}\tau }+\varepsilon .  \label{eq:46}
\end{eqnarray}%
Here, the parameters $H_{B,mn}$, $H_{C,mn}$, and $H_{D,mn}$ are
\begin{equation}
H_{C,mn} =\mu _{mn}^{\ast }C_{01,00}(0) +\chi _{mn}^{\ast }C_{10,00}(0)
-i\xi \nu _{mn}^{\ast },  \label{eq:47}
\end{equation}
and
\begin{equation}
H_{X,mn} =\mu _{mn}X_{00,01}(0) +\chi _{mn}X_{00,10}(0) +i\xi \nu _{mn}.
\label{eq:48}
\end{equation}
Here, the subscript $X$ can be either $B$ or $D$. With the same meaning as
in Eq.~(\ref{eq:39-1}), the parameters, e.g., $C_{01,00}(0)$, represent the
matrix elements of the operators $C(0)$, $B(0)$, and $D(0)$ in the truncated
basis in Eq.~(\ref{eq:25}). We note that the initial conditions
\begin{eqnarray}
B(0) &=& C(0)=\rho^{ss}\,a(0) ,  \label{eq:49} \\
D(0) &=& A(0)=\rho^{ss}\,a^{\dag }(0)  \label{eq:50}
\end{eqnarray}
are used when Eqs.~(\ref{eq:38}--\ref{eq:40}) are derived. The matrix
elements, e.g., $B_{00,01}(0)$, can also be obtained as
\begin{eqnarray}
B_{00,01}(0) &=&\rho _{00,00}^{ss},
\end{eqnarray}
using the quantum regression theorem with the matrix element $\rho
_{00,00}^{ss}$ of the reduced density matrix $\rho$.

By using Eq.~(\ref{eq:33}) and Eqs.~(\ref{eq:38}--\ref{eq:40}), the two-time
correlation function in Eq.~(\ref{correlation}) for the induced
electromotive force is given by
\begin{eqnarray}
\left\langle V(0) V(\tau) \right\rangle &\propto &A_{01,00}(\tau)
+B_{00,01}(\tau)  \notag \\
&&-C_{01,00}(\tau) -D_{00,01}(\tau).  \label{eq:56}
\end{eqnarray}%
Based on the above results, the two-time correlation function in Eq.~(\ref%
{eq:56}) is further simplified to
\begin{eqnarray}
\left\langle V(0) V(\tau) \right\rangle &\propto &\mu _{12}e^{-\lambda
_{1}\tau }+\mu _{21}e^{-\lambda _{2}\tau }  \notag \\
&&+f_{1}e^{-\lambda _{1}^{\ast }\tau }+f_{2}e^{-\lambda _{2}^{\ast }\tau }
\label{eq:57}
\end{eqnarray}%
with
\begin{eqnarray*}
f_{1} &=&\mu _{12}^{\ast }\rho _{01,01}^{ss}+\chi _{12}^{\ast }\rho
_{10,01}^{ss}, \\
f_{2} &=&\mu _{21}^{\ast }\rho _{01,01}^{ss}+\chi _{21}^{\ast }\rho
_{10,01}^{ss}.
\end{eqnarray*}

Then, replacing $\left\langle V(0) V(\tau) \right\rangle$ in Eq.~(\ref{eq:21}%
) by Eq.~(\ref{eq:57}), and integrating, the spectrum $S_{V}(\omega)$ in
Eq.~(\ref{eq:21}) can be expressed as
\begin{eqnarray}
S_{V}(\omega) &\approx &\frac{B^2l^2\Omega}{2M}\text{Re}\left[%
\int_{0}^{\infty }d\tau e^{i\omega \tau }\left( \mu _{12}e^{-\lambda
_{1}\tau }+\mu _{21}e^{-\lambda _{2}\tau }\right)\right]  \notag \\
&=&\frac{B^2l^2\Omega}{2M}\frac{ \Gamma _{1}\text{Re}( \mu _{12}) -\left(
\omega -\varphi _{1}\right) \text{Im}( \mu _{12}) }{\left( \omega -\varphi
_{1}\right) ^{2}+\Gamma _{1}^{2}}  \notag \\
&+&\frac{B^2l^2\Omega}{2M}\frac{\Gamma _{2}\text{Re}( \mu _{21}) -\left(
\omega -\varphi _{2}\right) \text{Im}( \mu _{21}) }{\left( \omega -\varphi
_{2}\right) ^{2}+\Gamma _{2}^{2}}.  \label{eq:58}
\end{eqnarray}%
In Eq.~(\ref{eq:58}), we have neglected the terms proportional to the
amplitudes $f_{1}$ and $f_{2}$. This because the ratios, e.g., $%
(f_{1}/\mu_{12})$, are proportional to $\xi ^{2}$, which is negligibly small
in the weak driving limit. In this case, we need only consider the two
leading terms, which are proportional to $\mu_{12}$ and $\mu_{21}$, as shown
in Eq.~(\ref{eq:58}).

\begin{figure}[th]
\includegraphics[bb=23 431 517 767,width=8cm]{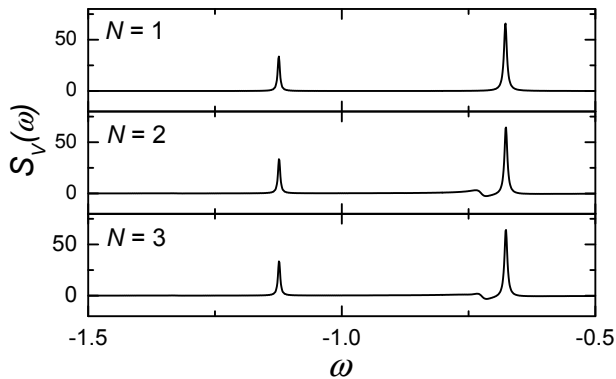}
\caption{The spectrum $S_{V}(\protect\omega) $ versus $\protect\omega$ for
three different values of the total number of excitations $N$, e.g., $%
N=1,\,2,\,3$. We take the values for other parameters as $\protect\delta=0.2
$, $g=0.2$, $\protect\xi =0.02$, $\protect\kappa =0.004$ and $\protect\gamma %
=0.004$. Here, all these parameters are in units of $1$ GHz.}
\label{number}
\end{figure}

\begin{figure}[th]
\includegraphics[bb=23 431 517 767, width=8cm]{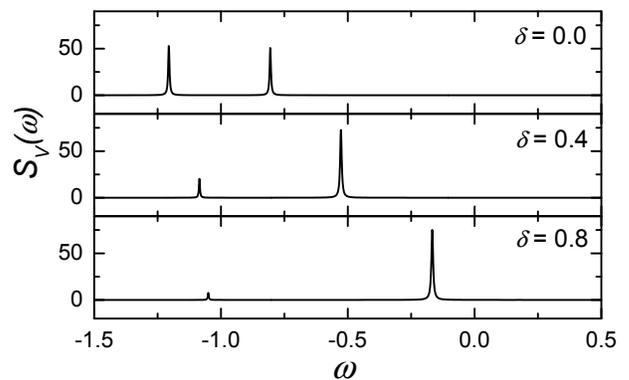}
\caption{The spectrum $S_{V}( \protect\omega) $ versus $\protect\omega$ for
three different values of the detuning $\protect\delta$, e.g., $\protect%
\delta=0,\, 0.4,\, 0.8$. The detuning $\protect\delta$ is defined in Eq.~(%
\protect\ref{delta}). The total excitation number $N$ is taken here as $N=1$%
. Other parameters and units are the same as in Fig.~\protect\ref{number}. }
\label{frequency}
\end{figure}

\begin{figure}[th]
\includegraphics[bb=23 431 517 767,width=8cm]{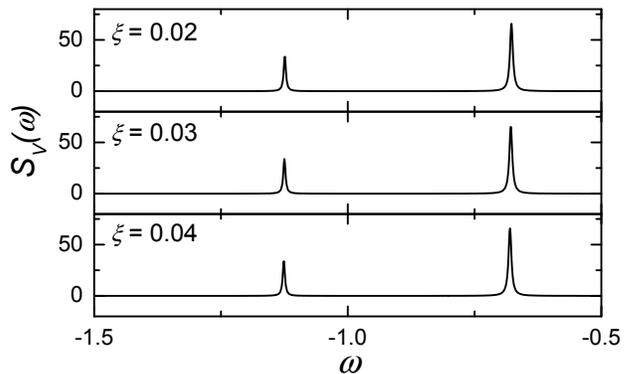}
\caption{The spectrum $S_{V}( \protect\omega) $ versus $\protect\omega$ for
three different values of the driving strength $\protect\xi$, e.g., $\protect%
\xi=0.02,\, 0.03,\, 0.04$. The coupling strength $\protect\xi \equiv \protect%
\xi _{\mathrm{NAMR-current}}$ is defined in Eq.~(\protect\ref{xi}). The
total excitation number $N$ is taken here as $N=1$. Other parameters and
units are the same as in Fig.~\protect\ref{number}.}
\label{drive}
\end{figure}

As shown in Figs.~\ref{number},\,\ref{frequency},\,\ref{drive}, there are
two dominant peaks in the $S_{V}(\omega)$ spectrum. The distance (splitting
frequency $\Delta \omega$) between these two peaks is
\begin{equation}  \label{splitting1}
\Delta \omega =\left( a^{2}+b^{2}\right) ^{\frac{1}{4}}\cos \left[ \frac{1}{2%
}\arctan \left( \frac{b}{a}\right) \right],
\end{equation}%
which is determined by the frequency detuning $\delta $ and the decay rates $%
\kappa $ and $\gamma $ for the NAMR and the charge qubit. Then, the
information of the coherent coupling between the charge qubit and the NAMR
can be obtained by Eq.~(\ref{splitting1}).

\subsection{Numerical results}

To test the validity of our analytical calculations obtained using both
quantum regression theorem and perturbation theory, we now study the
correlation spectrum $S_{V}(\omega)$ numerically.

In Fig.~\ref{number}, the spectrum $S_{V}(\omega)$ versus frequency $\omega$
is plotted with parameters $\delta=0.2$, $g=0.2$, $\xi=0.02$, $\kappa=0.004$%
, and $\gamma=0.004$, for different total excitation numbers $N$, e.g., $%
N=1,\,2,\,3$. Here, we take $1$ GHz as the unit for all these parameters.

Figure~\ref{number} shows that: (i) there are two prominent peaks, which
means that the approximation in Eq.~(\ref{eq:58}) is valid in the weak
driving limit; (ii) the increase of the total number $N$ of the excitation
does not obviously change the heights and the splitting frequency $\Delta
\omega$ of the two leading peaks. Therefore, Fig.~\ref{number} verifies that
the approximation with the truncated basis ($N=1$) is valid in the limit of
weak driving ($\xi \ll g$).

Now in the truncated basis ($N=1$), let us demonstrate the effects of the
frequency detuning $\delta $ [see Eq.~(\ref{delta})] and the driving
strength $\xi $ [see Eq.~(\ref{xi})] on the splitting frequency $\Delta
\omega $. In Fig.~\ref{frequency}, the spectrum $S_{V}(\omega)$ versus
frequency $\omega$ is plotted for different detunings $\delta$, and the same
other parameters as in Fig.~\ref{number}. Figure~\ref{frequency} shows that
the detuning $\delta$ affects both the heights and the frequency splitting $%
\Delta \omega$ of the two peaks. A larger $\delta$ corresponds to a
larger frequency splitting $\Delta \omega$. When $\delta$ is
increased, the heights of the two peaks change
\emph{asymmetrically}, i.e., one peak becomes higher than another
one when increasing the detuning $\delta$. (Similar results were
found in Ref.~\cite{Wei06}). This means that our probing approach
does not work well when the frequency
detuning ($\delta=0.8$) is much larger than the interaction strength ($g=0.2$%
). In practice, we should assume that the frequency detuning
$\delta$ and the interaction strength $g$ is of the same order.
Also, Fig.~\ref{frequency} shows that the detuning $\delta$ does not
affect the number of peaks, which means that the approximation in
Eq.~(\ref{eq:58}) is valid in the weak driving limit.

Similarly, in Fig.~\ref{drive}, the spectrum $S_{V}(\omega)$ versus
frequency $\omega$ is plotted for different driving strengths $\xi$
and the same other parameters as in Fig.~\ref{number}, when the
total excitation number $N=1$. Figure~\ref{drive} shows that a weak
driving strength $\xi$ does not significantly affect the frequency
splitting $\Delta \omega$ of the two peaks, and also it does not
affect the number of peaks. Moreover, changing the driving strength
$\xi$ does not affect the heights of both peaks.

\section{conclusions}

We have proposed an ``intrinsic probing'' approach to demonstrate the
coherent coupling between a driven metallic NAMR and a charge qubit. This
metallic NAMR can act as an intrinsic detector when a weak driving current
passes through it. Using the quantum regression theorem and perturbation
theory, we have calculated the correlation spectrum of the electromotive
force between two ends of the NAMR. This spectrum can be used to demonstrate
QED analogues in the NAMR-qubit system, e.g., the vacuum Rabi splitting
related to the coherent coupling strength of the charge qubit to the NAMR.
The numerical calculations confirm the validity of the analytical results.

In our proposal, no additional measurement instruments needs to be
integrated, in contrast to the proposal in~\cite{Wei06}. The NAMR acts as
both the cavity and the detector. Therefore, it is easier to be fabricated.
Our proposal can also be generalized to the case where many qubits are
coupled to a NAMR. In this case, the information of many qubits can also be
readout via the spectrum of the electromotive force. Recent experiments~\cite%
{Litf08} indicate that our proposal is experimentally realizable.

\begin{acknowledgements}

We acknowledge the support of the NSFC Grant Nos. 10547101,
10604002, the National Fundamental Research Program of China Grant
No. 2006CB921200. FN acknowledges partial support from the National
Security Agency (NSA), Laboratory Physical Science (LPS), Army
Research Office (ARO), National Science Foundation (NSF) grant No.
EIA-0130383, JSPS-RFBR 06-02-91200, and Core-to-Core (CTC) program
supported by the Japan Society for Promotion of Science (JSPS). We
also thank N. Zhao for discussions.

\end{acknowledgements}

\end{document}